\documentclass[twocolumn]{aastex62}

\usepackage[utf8]{inputenc}
\usepackage{graphicx}
\usepackage{amsmath}
\graphicspath{{Figures/}}

\usepackage{gensymb}

\newcommand{\ucb}{\affiliation{Physics Department, University of California, Berkeley, CA 94720-7300, USA}}
\newcommand{\ssl}{\affiliation{Space Sciences Laboratory, University of California, Berkeley, CA 94720-7450, USA}}
\newcommand{\qml}{\affiliation{School of Physics and Astronomy, Queen Mary University of London, London E1 4NS, UK}}
\newcommand{\unh}{\affiliation{Department of Physics and Astronomy, University of New Hampshire, Durham, NH 03824, USA}}

\newcommand\Alfvenic{Alfv\'enic}

\begin{document}

\title{Cross Helicity Reversals In Magnetic Switchbacks}

\begin{abstract}
We consider 2D joint distributions of normalised residual energy $\sigma_r(s,t)$ and cross helicity $\sigma_c(s,t)$ during one day of Parker Solar Probe's (PSP's) first encounter as a function of wavelet scale $s$. The broad features of the distributions are similar to previous observations made by HELIOS in slow solar wind, namely well correlated and fairly Alfv\'enic, except for a population with negative cross helicity which is seen at shorter wavelet scales. We show that this population is due to the presence of magnetic switchbacks, brief periods where the magnetic field polarity reverses. Such switchbacks have been observed before, both in HELIOS data and in Ulysses data in the polar solar wind. Their abundance and short timescales as seen by PSP in its first encounter is a new observation, and their precise origin is still unknown. By analysing these MHD invariants as a function of wavelet scale we show that MHD waves do indeed follow the local mean magnetic field through switchbacks, with net Elsasser flux propagating inward during the field reversal, and that they therefore must be local kinks in the magnetic field and not due to small regions of opposite polarity on the surface of the Sun. Such observations are important to keep in mind as computing cross helicity without taking into account the effect of switchbacks may result in spurious underestimation of $\sigma_c$ as PSP gets closer to the Sun in later orbits.
\end{abstract}

\author[0000-0001-6077-4145]{Michael D. McManus}\ucb\ssl
\correspondingauthor{Michael D. McManus}
\email{mdmcmanus@berkeley.edu}
\author[0000-0002-4625-3332]{Trevor A. Bowen}\ssl
\author[0000-0001-9202-1340]{Alfred Mallet}\ssl
\author[0000-0003-4529-3620]{Christopher H. K. Chen}\qml
\author[0000-0003-4177-3328]{Benjamin D. G. Chandran}\unh
\author[0000-0002-1989-3596]{Stuart D. Bale}\ssl
\author[0000-0003-4529-3620]{Davin E. Larson}\ssl

\author[0000-0002-4401-0943]{Thierry {Dudok de Wit}}
\affiliation{LPC2E, CNRS and University of Orl\'eans, Orl\'eans, France}

\author[0000-0002-7077-930X]{J. C. Kasper}
\affiliation{Climate and Space Sciences and Engineering, University of Michigan, Ann Arbor, MI 48109, USA}
\affiliation{Smithsonian Astrophysical Observatory, Cambridge, MA 02138 USA}

\author[0000-0002-7728-0085]{Michael Stevens}
\affiliation{Smithsonian Astrophysical Observatory, Cambridge, MA 02138 USA}

\author[0000-0002-7287-5098]{Phyllis Whittlesey}
\affiliation{Space Sciences Laboratory, University of California, Berkeley, CA 94720-7450, USA}

\author[0000-0002-0396-0547]{Roberto Livi}
\affiliation{Space Sciences Laboratory, University of California, Berkeley, CA 94720-7450, USA}

\author[0000-0001-6095-2490]{Kelly E. Korreck}
\affiliation{Smithsonian Astrophysical Observatory, Cambridge, MA 02138 USA}

\author[0000-0003-0420-3633]{Keith Goetz}
\affiliation{School of Physics and Astronomy, University of Minnesota, Minneapolis, MN 55455, USA}

\author[0000-0002-6938-0166]{Peter R. Harvey}
\affiliation{Space Sciences Laboratory, University of California, Berkeley, CA 94720-7450, USA}

\author[0000-0002-1573-7457]{Marc Pulupa}
\affiliation{Space Sciences Laboratory, University of California, Berkeley, CA 94720-7450, USA}

\author[0000-0003-3112-4201]{Robert J. MacDowall}
\affiliation{Solar System Exploration Division, NASA/Goddard Space Flight Center, Greenbelt, MD, 20771}

\author[0000-0003-1191-1558]{David M. Malaspina}
\affil{Laboratory for Atmospheric and Space Physics, University of Colorado, Boulder, CO 80303, USA}

\author[0000-0002-3520-4041]{Anthony W. Case}
\affiliation{Smithsonian Astrophysical Observatory, Cambridge, MA 02138 USA.}

\author[0000-0002-0675-7907]{J. W. Bonnell}
\affil{Space Sciences Laboratory, University of California, Berkeley, CA 94720-7450, USA}


\section{Introduction}


Parker Solar Probe (PSP) \citep{fox2016solar} was launched in August 2018 with the aim of shedding light on the plasma and magnetic field environments of the inner heliosphere and the longstanding problem of coronal heating. It completed its first of a series of 24 encounters on November 11th 2018, during which at perihelion it was a distance of $35R_S$ from the Sun.

One of the more notable observations reported from the first encounter has been the preponderance of so called magnetic “switchbacks”, large traversals of the mainly radial magnetic field, often temporarily reversing the sense of the field. Prior to Parker Solar Probe, magnetic switchbacks had been observed both in near-Sun (0.3AU) HELIOS data \citep{horbury2018short}, and over the solar poles by Ulysses \citep{balogh1999heliospheric}. Both studies involved fast solar wind streams. After reprocessing HELIOS data, \cite{horbury2018short} found that large velocity spikes are ubiquitous in near-Sun fast solar wind, occurring about $5\%$ of the time and with magnitudes of order $0.5v_A$ above the background solar wind speed. The velocity spikes they observed were almost always positive speed enhancements, were highly \Alfvenic{} in all three components (thus by necessity accompanied by large magnetic field traversals), and showed no statistically meaningful difference in plasma parameters inside versus outside the spikes (making it unlikely that the observed field geometry was due to HELIOS crossing large coronal loops). The authors speculated that they may be the \Alfvenic{} fluctuations that travel ahead of jets generated by reconnection events in the corona \citep{karpen2017reconnection,uritsky2017reconnection}, and are thus signatures of transient events at the Sun’s surface that have survived to relatively large distances. 
The spikes or switchbacks seen by PSP in its first two encounters are qualitatively different than these in two ways; they are shorter in timescale (presumably due to being at smaller heliocentric distances and having better measurement cadences able to resolve sharper spikes), and they are the first direct observation of them in slow as opposed to fast solar wind, marking them to be a universal feature of the solar wind. 

Earlier work by \cite{balogh1999heliospheric} reported magnetic field inversions at high heliographic latitudes that lasted on the order of several hours, and used cross helicity as a sensor of wave propagation direction to deduce that the inversions they saw were not intrinsically different magnetic sectors but rather due to fold-like structures in the magnetic field. In this work we use wavelet representations of the dimensionless MHD transport ratios cross helicity, $\sigma_c$, and residual energy, $\sigma_r$, in a similar way to probe the geometry of the short timescale magnetic switchbacks seen by PSP in encounter 1 over a wide range of scales. We deduce that they too are due to localised folds in the magnetic field and not regions of different magnetic polarity.  

Several other sensors can be used to elucidate local magnetic field topology. Electron strahl pitch angle distributions, as measured by the SPAN instrument on PSP \citep{whittlesey2019span, livi2019span}, are used by  \cite{whittlesey2019strahl} to follow the magnetic field through switchbacks.   
\cite{neugebauer2013double} showed that the relative proton core-beam drift becomes negative (that is, the beam appears to be moving more slowly than the core in the spacecraft frame), whenever the local field switches back on itself, and \cite{yamauchi2004differential} used the alpha-proton differential velocity to show the same thing within the context of pressure balance structures. Our technique has the advantage of being somewhat less complex than these methods, requiring less detailed analysis of the particle distribution functions (only the perturbed bulk velocity moments are needed). 

This clear dependence of plasma properties on the local magnetic field is reflected in the plasma turbulence as well. Turbulent power is concentrated at near perpendicular angles $\theta_{BV}$ between the magnetic field and flow direction, and the magnetic field spectral index is a smoothly increasing function of $\theta_{BV}$ 
\citep{horbury2008anisotropic,podesta2009dependence,chen2011anisotropy}. This dependence of spectral index on $\theta_{BV}$ was only revealed when sufficient care was used to examine the mean field at small enough (i.e. localised) scales, via a wavelet method.

Throughout the solar wind we see \Alfvenic{} turbulence, and there are numerous models of how this turbulence behaves both at 1AU \citep{boldyrev2005spectrum,mallet2016statistical} and in the inner heliosphere \citep{velli1989turbulent,chandran2019reflection,perez2013direct}. The relationship between $\sigma_c$ and $\sigma_r$, as useful invariants to characterise the state of the MHD turbulence, has been well studied (\cite{bruno2007magnetically}, \cite{bruno2005solar} and references therein). Fast wind at short heliocentric distances is very \Alfvenic{} and equipartitioned ($\sigma_c \sim 1, \sigma_r \sim 0$), but a second population with $\sigma_c \sim 0, \sigma_r \sim -1$ appears as heliocentric distance increases, representing the presence of intermittent magnetic structures. The importance of negative residual energy and intermittency and how it causes the magnetic field spectrum to steepen was highlighted in \cite{bowen2018impact} and \cite{chen2013residual}. Slow wind does not show such marked radial evolution, with broader $(\sigma_c,\sigma_r)$ distributions in general. 

In section 2 we outline the data set and methods used, section 3 contains results and discussion, and we briefly summarise the conclusions in section 4. 

\begin{figure}
    \centering
    \includegraphics[scale=0.35]{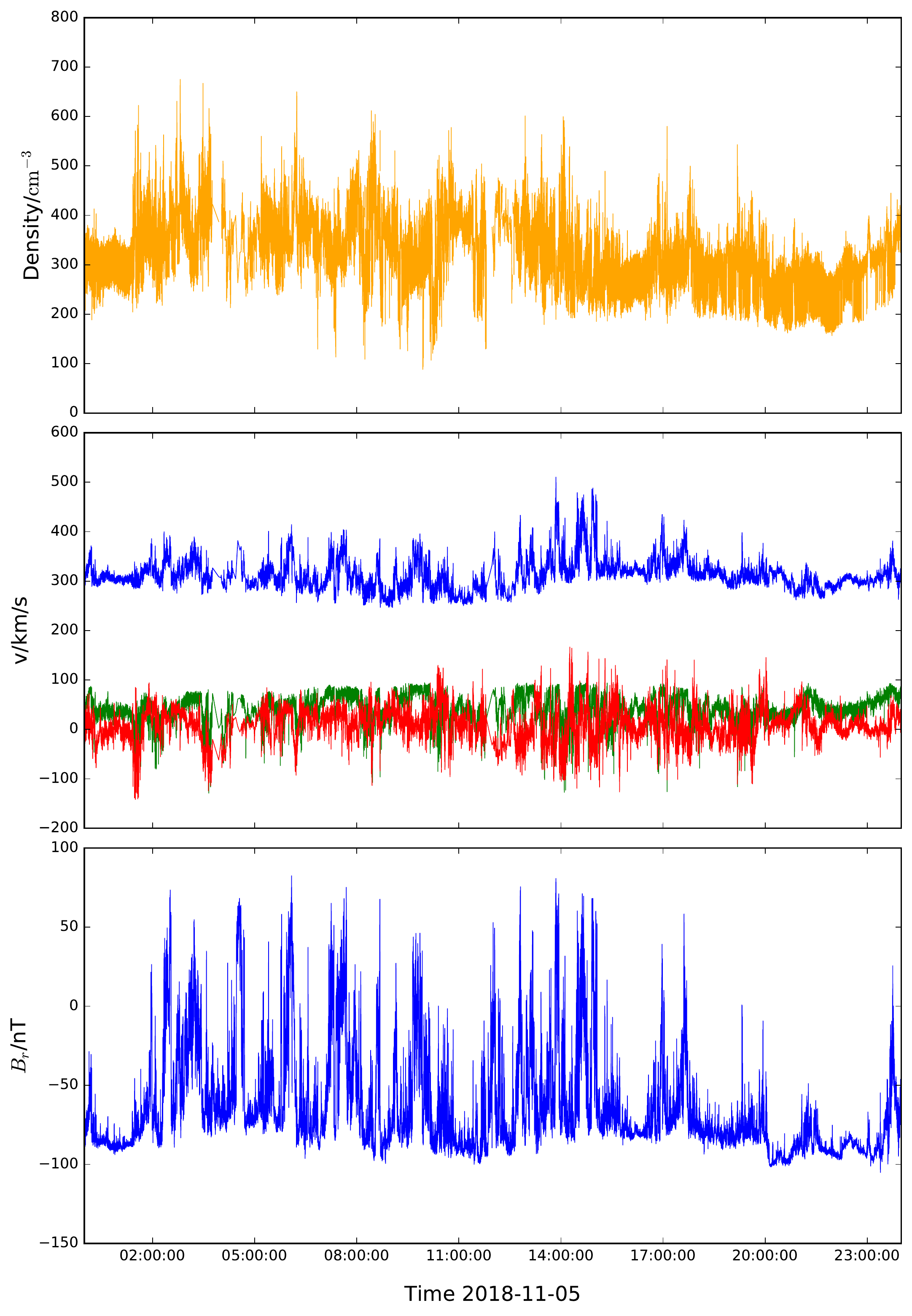}
    \caption{Time series of the encounter 1 interval. Top panel shows proton density, middle panel shows proton velocity moments in RTN coordinates from SPC (blue, green, red being radial, tangential, normal respectively), and the bottom panel shows radial component of the magnetic field.}
    \label{fig:enc_interval}
\end{figure}

\section{Data and Methods}
We use particle measurements of proton density $\rho$ and velocity $\mathbf{v}$ made by PSP's onboard Faraday cup, Solar Probe Cup (SPC) \citep{kasper2016solar}, and magnetic field measurements made by the FIELDS fluxgate magnetometer \citep{bale2016fields}. We consider a 1 day interval from encounter 1, Nov 5th 2018. 
The encounter 1 measurement cadence for SPC proton moments is approximately 0.87s, while the magnetometer measurement frequency was approximately 293Hz. The magnetometer data was downsampled to match SPC's measurement cadence, and an approximately 2.6s timing offset corrected for. Large unphysical spikes were also removed, and any data gaps linearly interpolated over. Figure 1 shows particle and magnetic field data for this interval. 

Throughout the analysis we make use of wavelet transform representations of various quantities. A wavelet transform of a discrete time series $x(t_i)$ is defined as \citep{torrence1998practical,addison2017illustrated}
\begin{equation}
    W(s,t) = \sum_{i=0}^{N-1} x(t_i) \psi \left(\frac{t_i - t}{s}  \right)
    \label{eq:cwt}
\end{equation}
where $W(s,t)$ is the wavelet coefficient at scale $s$ and time $t$, $\psi(t,s)$ the wavelet function and $\{t_i\}$ the set of measurement times. We use a Morlet wavelet \citep{farge1992wavelet} as our wavelet function (written here unnormalized), 
\begin{equation}
    \psi(t) = \pi^{\frac{1}{4}} e^{-\frac{1}{2}t^2}e^{i\sigma t},
\end{equation}
where $\sigma$ is an adjustable parameter taken here to be $6$ that represents the frequency of the wavelet. We convert from dilation scale $s$ to physical (spacecraft) frequency $f$ using
\begin{equation}
    f = \frac{\sigma}{2\pi \Delta t s}
\end{equation}
where $\Delta t$ is the measurement cadence. In this work we use 24 logarithmically spaced wavelet scales $s$, from $s_\text{min} = 2$ to $s_\text{max} = 5792.62$.

First, we compute a scale and time dependent local mean magnetic field $\mathbf{B}_0(s,t)$ as
\begin{equation}
    \mathbf{B}_0(s,t) = \sum_{i=0}^{N-1} \mathbf{B}(t_i) \left \lvert \psi \left(\frac{t_i - t}{s} \right)\right \rvert
    \label{eq:mean_field}
\end{equation}
\citep{horbury2008anisotropic,podesta2009dependence}. This convolution of $\mathbf{B}(t)$ with $\lvert \psi \rvert$ can be intuitively understood as a smoothing of $\mathbf{B}(t)$ over a window whose size is determined by the width of the Morlet wavelet's Gaussian envelope, $|\psi|$, which in turn is set by the scale length $s$. We then apply the wavelet transform \ref{eq:cwt} to the time series $\mathbf{v}(t)$ and $\mathbf{b}(t)$, which gives us the scale and time dependent fluctuations $\delta \mathbf{v}(s,t)$ and $\delta \mathbf{b}(s,t)$ (since the wavelet transform has no zero frequency component). With \ref{eq:mean_field} a local parallel field direction is defined, from which we can calculate the wavelet representations of the perpendicular fluctuations $\delta \mathbf{v}_{\perp}(s,t)$ and $\delta \mathbf{b}_{\perp}(s,t)$, and the perpendicular Elsasser variables
\begin{equation}
    \delta \mathbf{z}_\perp^{\pm} (s,t) = \delta \mathbf{v}_\perp (s,t) \pm \delta \mathbf{b}_\perp (s,t)
    \label{eq:elsasser}.
\end{equation}
Here $\delta \mathbf{b}_{\perp}(s,t)$ is measured in Alfv\'en units. To convert we use a scale and time dependent density $\rho(s,t)$ computed using equation \ref{eq:mean_field} applied to density,
\begin{equation}
    \rho(s,t) = \sum_{i=0}^{N-1} \rho(t_i) \left \lvert \psi \left(\frac{t_i - t}{s} \right)\right \rvert.
\end{equation}
\begin{figure}
    \centering
    \includegraphics[scale=0.4]{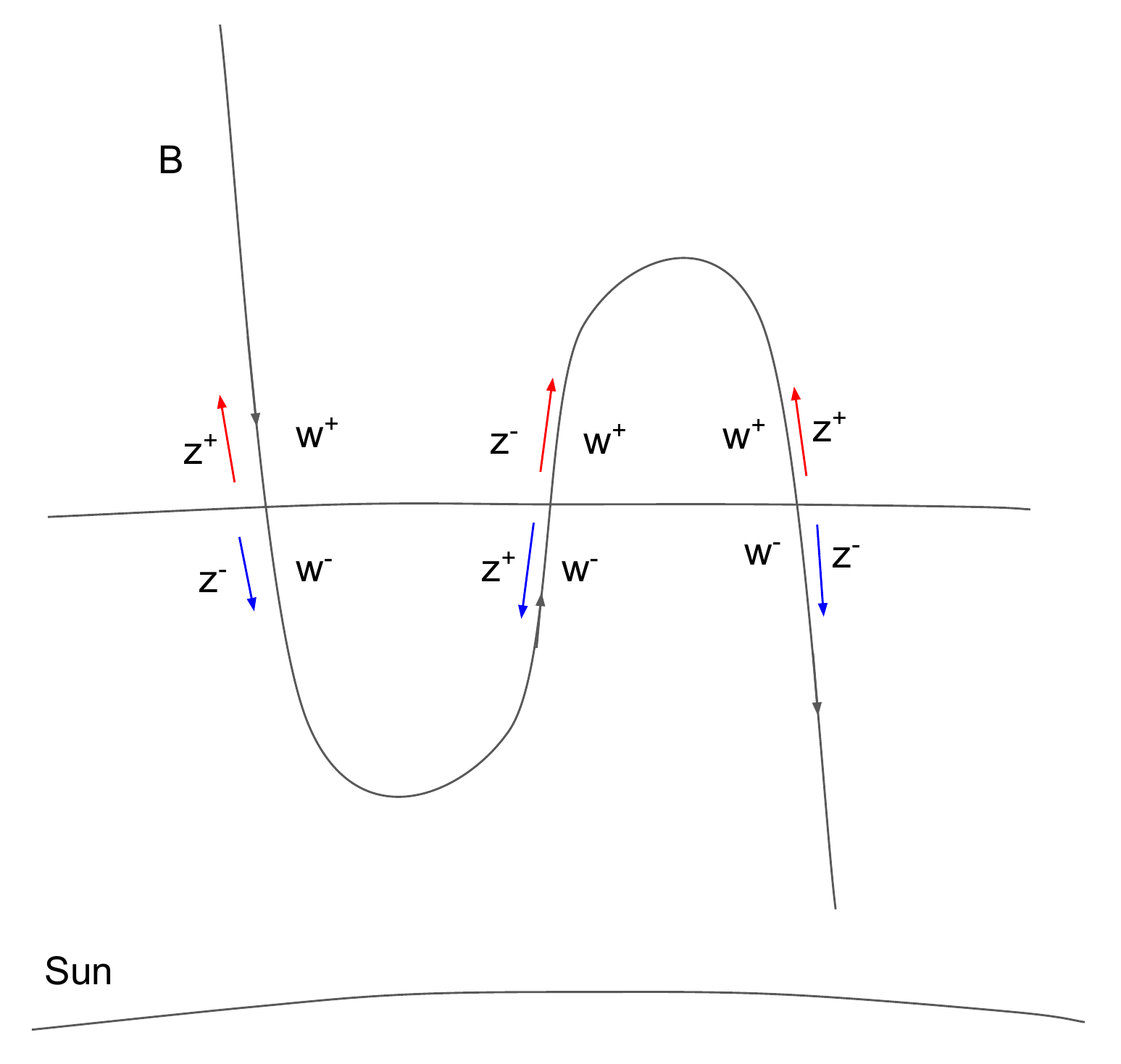}
    \caption{Schematic of a magnetic switchback, showing the redefinition of $\delta \mathbf{w}_\perp^{\pm}$ in terms of $\delta \mathbf{z}_\perp^{\pm}$ when $B_r$ changes sign.}
    \label{fig:cartoon}
\end{figure}

\begin{figure*}
    \includegraphics[width=\textwidth]{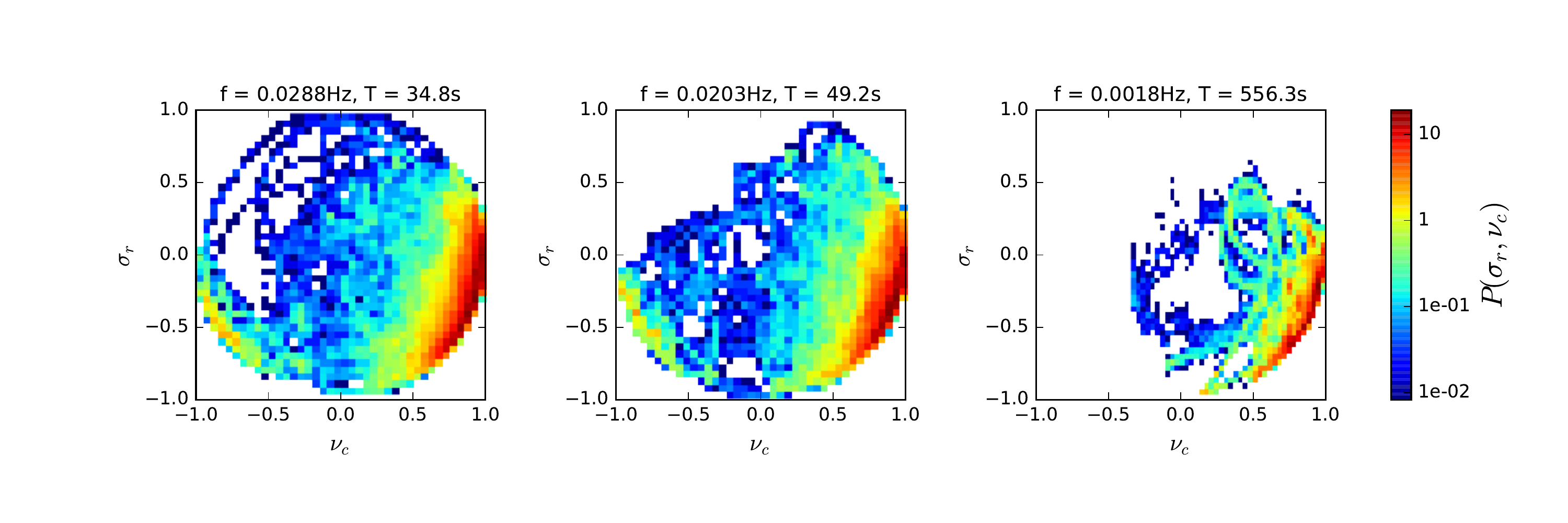}
    \caption{Joint probability distribution histograms of $\sigma_r$ vs $\nu_c$ for one day of encounter 1 (2018-11-05) at three different wavelet scales, from left to right: $T=35$s, $T=49$s, $T=556$s.}
    \label{fig:circles_all}
\end{figure*}
It is usual in the solar wind literature \citep{bavassano1998cross,bavassano2006distribution, roberts1987nature} to define $\delta \mathbf{z}_\perp^{\pm}$ in such a way that $\delta \mathbf{z}_\perp^{+}$ and $\delta \mathbf{z}_\perp^{-}$ always refer to outward and inward going waves respectively, regardless of the direction of the background magnetic field. Since $\mathbf{z}_\perp^{+}$ and $\mathbf{z}_\perp^{-}$ wave packets travel anti-parallel/parallel to $\mathbf{B}_0$ respectively, a scheme of magnetic ``rectification" is usually employed, flipping $\mathbf{B}_0$ as necessary. While this is useful when dealing with large scale magnetic sectors of different polarity, it will be much clearer in the following discussion to leave the definition of $\delta \mathbf{z}_\perp^{\pm}$ as is in equation \ref{eq:elsasser}, and define two new variables, $\delta \mathbf{w}_\perp^{\pm}$, to represent strictly outgoing ($+$) and ingoing ($-$) waves respectively: 
\begin{align}
    \delta \mathbf{w}_\perp^{\pm}(s,t) &= 
    \begin{cases}
        \delta \mathbf{z}_\perp^{\pm}(s,t) & \text{if $\text{sgn}\left(B_{0r}(s,t)\right) = -1$} \\
        \delta \mathbf{z}_\perp^{\mp}(s,t) & \text{if $\text{sgn}\left(B_{0r}(s,t)\right) = 1$}   
    \end{cases}
\end{align}
where $B_{0r}(s,t)$ is the radial component (in RTN coordinates) of the scale dependent mean magnetic field defined in equation \ref{eq:mean_field}. Physically this is equivalent to the usual method of rectifying the field. The cartoon in Figure \ref{fig:cartoon} illustrates these definitions for a situation where PSP observes a field polarity reversal in an overall radially inward background field. For illustration we have drawn this as an S-shaped bend, but a priori the exact field geometry is unknown. 

To define switchback times, we first compute the time average over the entire interval of the radial component of the background magnetic field, $\langle B_{0r}(s_\text{max},t) \rangle_t$, at the largest wavelet scale $s_\text{max}$. We define the overall sense of the background magnetic field to be $\eta \equiv \text{sgn}(\langle B_{0r}(s_\text{max},t) \rangle_t)$. At each wavelet scale then, we can define a magnetic inversion or switchback to be when $B_{0r}(s,t)$ changes sign, relative to this largest scale background magnetic field direction. In other words, when $B_{0r}(s,t) = - \eta$.

With these definitions in hand we can compute the normalised residual energy
\begin{align}\label{eq:sigma_r}
    \sigma_r(s,t) &= \frac{\lvert \delta \mathbf{v}_\perp(s,t) \rvert^2 - \lvert \delta \mathbf{b}_\perp(s,t) \rvert^2 }{\lvert \delta \mathbf{v}_\perp(s,t) \rvert^2 + \lvert \delta \mathbf{b}_\perp(s,t) \rvert^2} \\ 
    &= \frac{2 \delta \mathbf{z}^{+}_\perp(s,t) \cdot \delta \mathbf{z}^{-}_\perp(s,t)}{\lvert \delta \mathbf{z}^{+}_\perp(s,t) \rvert^2 + 
    \lvert \delta \mathbf{z}^{-}_\perp(s,t) \rvert^2},
\end{align}
which represents the imbalance between kinetic and magnetic fluctuations, or equivalently the alignment between the two Elsasser variables, and normalised cross helicity,
\begin{align}\label{eq:sigma_c}
    \sigma_c(s,t) &= \frac{2 \delta \mathbf{v}_\perp(s,t) \cdot \delta \mathbf{b}_\perp(s,t)}{\lvert \delta \mathbf{v}_\perp(s,t) \rvert^2 + \lvert \delta \mathbf{b}_\perp(s,t) \rvert^2} \\
    &= \frac{\lvert \delta \mathbf{z}^{+}_\perp(s,t) \rvert^2 - 
    \lvert \delta \mathbf{z}^{-}_\perp(s,t) \rvert^2}{\lvert \delta \mathbf{z}^{+}_\perp(s,t) \rvert^2 + 
    \lvert \delta \mathbf{z}^{-}_\perp(s,t) \rvert^2},
\end{align}
representing the alignment between velocity and magnetic field fluctuations, or the imbalance between the flux of $\delta \mathbf{z}^{+}_\perp$ and $\delta \mathbf{z}^{-}_\perp$. By analogy we have the ``rectified" cross helicity, constructed using $\delta \mathbf{w}_\perp^{\pm}$, which we will denote $\nu_c$:
\begin{align}\label{eq:nu_c}
    \nu_c(s,t) &= \frac{\lvert \delta \mathbf{w}^{+}_\perp(s,t) \rvert^2 - 
    \lvert \delta \mathbf{w}^{-}_\perp(s,t) \rvert^2}{\lvert \delta \mathbf{w}^{+}_\perp(s,t) \rvert^2 + 
    \lvert \delta \mathbf{w}^{-}_\perp(s,t) \rvert^2}
\end{align}
(rectification does not affect $\sigma_r$). $\nu_c$ is therefore a sensor of ingoing vs outgoing Elsasser flux, with respect to the radial direction $\hat{\mathbf{r}}$, regardless of the direction of the mean magnetic field. 
It is helpful to think of $\sigma_c$ as the fractional excess of fluctuations propagating anti-parallel to $\mathbf{B_0}$, and $\nu_c$ as the fractional excess of fluctuations propagating away from the Sun. 

Equations \ref{eq:sigma_r}, \ref{eq:sigma_c}, and \ref{eq:nu_c} impose the geometric constraint that
\begin{align}
    \sigma_c^2 + \sigma_r^2 &\le 1 \\
    \nu_c^2 + \sigma_r^2 &\le 1,
\end{align}
i.e. points in $(\sigma_c,\sigma_r)$ and $(\nu_c,\sigma_r)$ space are constrained to lie within a circle of radius 1. For a purely Alfv\'enic fluctuation, $\sigma_r = 0$ and $\nu_c = \pm 1$, with $+$ representing an outgoing wave and $-$ an ingoing one. Values of $|\nu_c| < 1$ represent either mixtures of ingoing and outgoing modes or mixtures of Alfv\'enic and non-Alfv\'enic fluctuations, two situations which cannot be distinguished by examining $\nu_c$ alone. 

Finally, we define the inward and outward going Elsasser fluxes
\begin{align}
    e^+ &= \lvert \delta \mathbf{w}^{+}_\perp(s,t) \rvert^2 \\
    e^- &= \lvert \delta \mathbf{w}^{-}_\perp(s,t) \rvert^2.
\end{align}

\section{Results and Discussion}
\begin{figure*}
    \includegraphics[width=\textwidth]{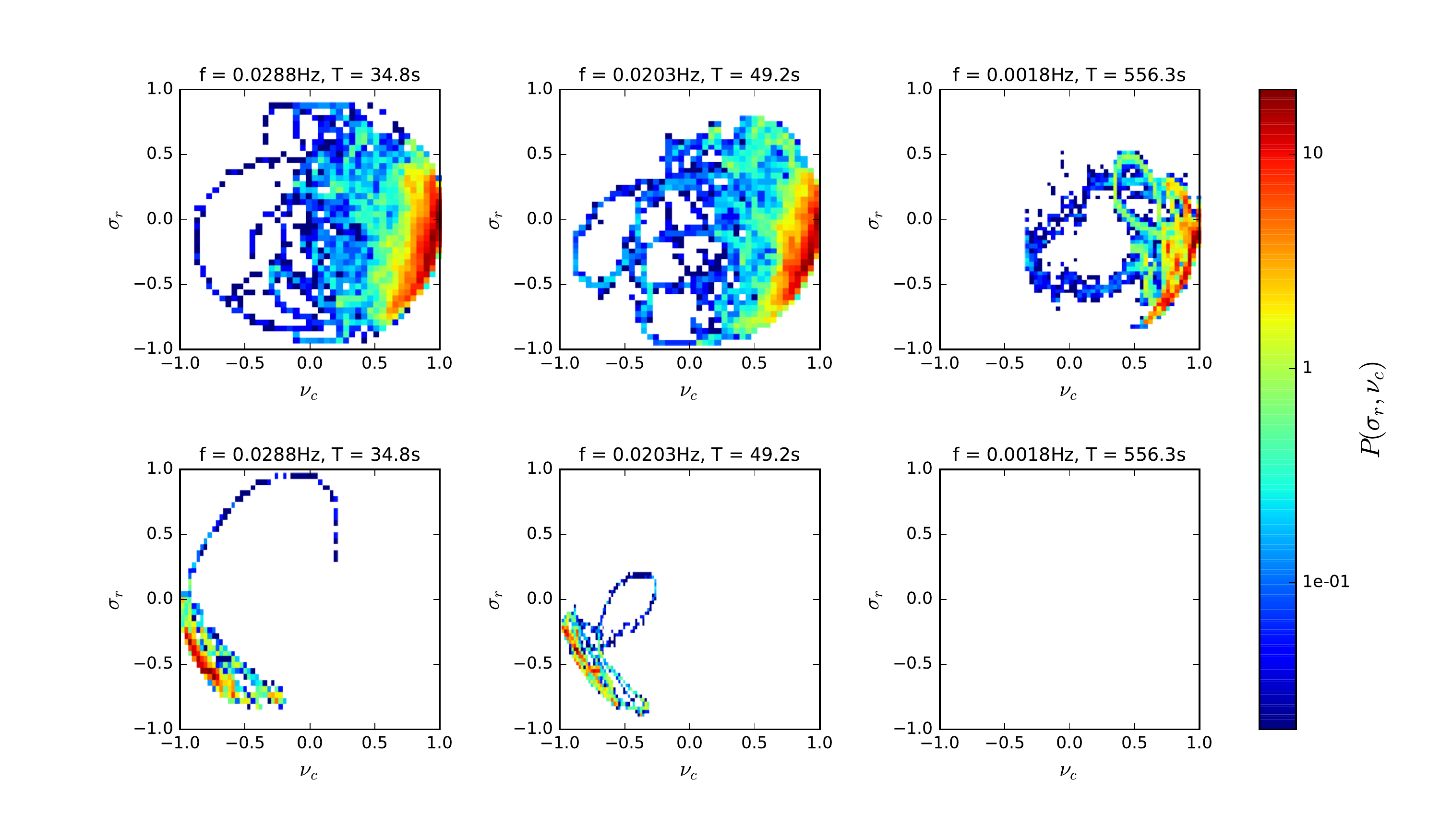}
    \caption{Joint probability distribution histograms of $\sigma_r$ and $\nu_c$ at three different wavelet scales, from left to right: $T=35$s, $T=49$s, $T=556$s, divided by $\theta_{Br}$. Top row: only those times on when $\theta_{Br} > 160\degree$, corresponding to a mainly radial field. Bottom row: only times when $\theta_{Br} < 90\degree$, when the radial magnetic field has locally reversed.}
    \label{fig:circles_angle}
\end{figure*}
Figure \ref{fig:enc_interval} shows the day-long interval during encounter 1 used in this analysis. The solar wind speed is relatively low, $v_{\text{sw}} \approx 330$km/s (throughout encounter 1 PSP was connected mainly to the same equatorial coronal hole \citep{badman2019PFSS}), and the radial distance is $R = 0.17AU$. The bottom panel shows the radial component of the magnetic field. The overall sense of the magnetic field is radially inwards, but a forest of narrow, spiky switchbacks where $B_r$ becomes positive are clearly visible. 
In Figure \ref{fig:circles_all} we plot joint histograms of $\sigma_r$ vs $\nu_c$ at three different wavelet scales, two short ones (35s and 50s), and one relatively longer one (560s). These frequencies are all well above the SPC velocity moment noise floor, which in this case corresponds to a frequency $f \approx 0.12$Hz or scale $T \approx 8.3$s. All three histograms are strongly peaked in the bottom right quadrant, near the edge of the limiting circle, with maxima around $\nu_c \sim 0.9, \sigma_r \sim -0.3$, indicating highly aligned \citep{wicks2013correlations} and fairly Alfv\'enic fluctuations. Of interest is the clear signal of a ``second population" at the two smaller scales, seen as a peak in the lower left quadrant with fewer counts and similar values of $\sigma_r \sim -0.3$ but with negative values of $\nu_c \sim -0.9$. No such population is seen at the longer $560$s time scale (and indeed at any wavelet scale longer than this).

The physical origin of the negative cross helicity population can be easily understood. In Figure \ref{fig:circles_angle} we divide up the data according to $\theta_{Br}$, the angle between the local magnetic field $\mathbf{B}_0(s,t)$ and the radial direction. The top row is histograms of $\sigma_r$ vs $\nu_c$ but only including times for which $\theta_{Br} > 160 \degree$ - a mainly radial field. The second row includes only times when $\theta_{Br} < 90\degree$, when the magnetic field has undergone a switchback.

The negative helicity population has clearly separated and is identifiable precisely with switchback intervals. This suggests that inside switchbacks MHD waves do indeed follow the \emph{local} magnetic field - the negative cross helicity values represent what was once majority outgoing waves becoming predominantly inward propagating inside a switchback (refer again to figure \ref{fig:cartoon}). This also implies that magnetic switchbacks are local kinks in the magnetic field and not due to small regions of opposite polarity at the surface of the Sun (in agreement with the conclusions in \cite{whittlesey2019strahl}). It is worth remarking here that by ``inward propagating" we mean relative to the plasma frame, not the spacecraft frame, since the Alfv\'en velocity is much smaller than the solar wind speed. 
\begin{figure*}
    \centering
    \includegraphics[width=\textwidth]{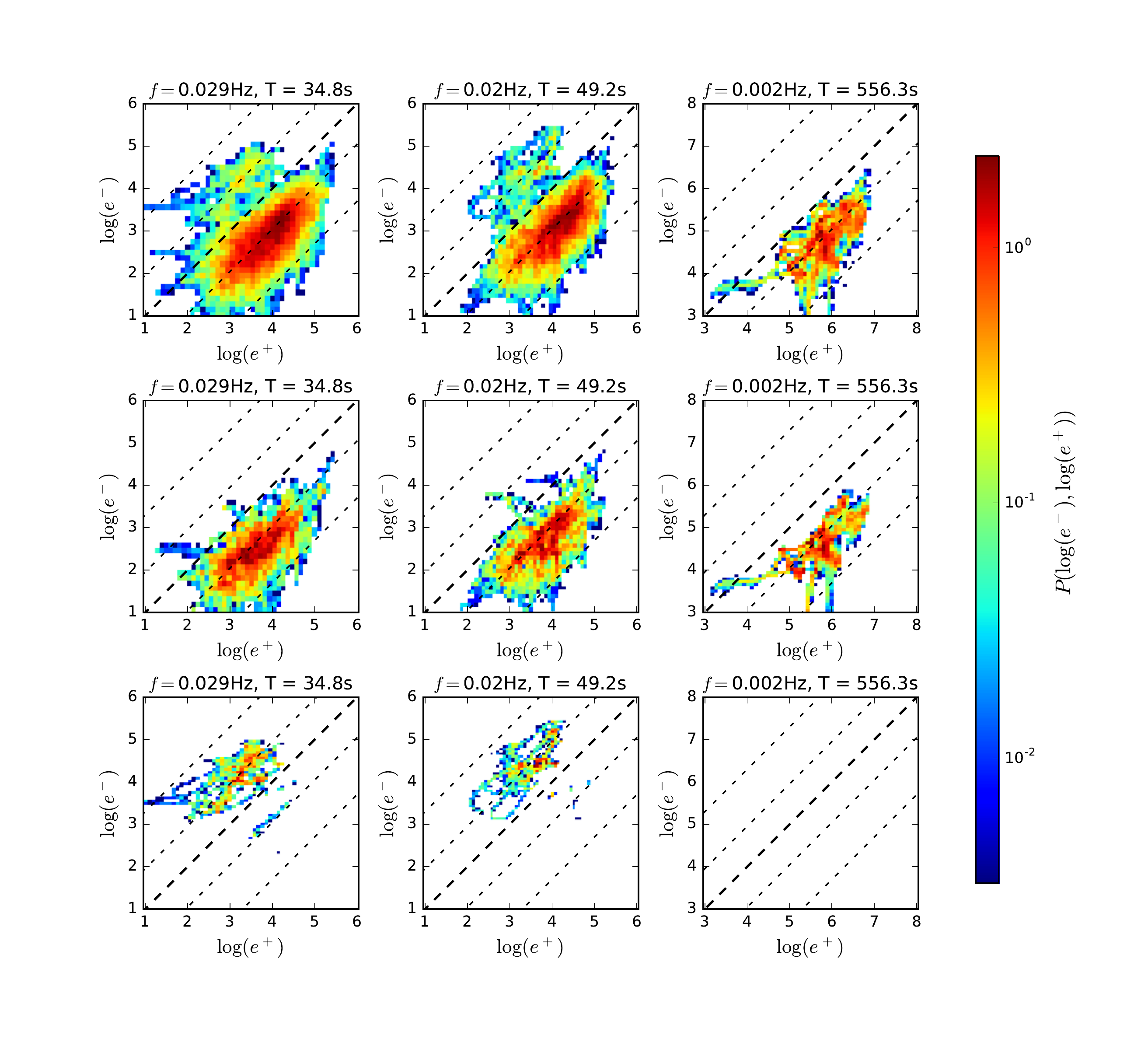}
    \caption{Joint probability distribution histograms of Elsasser power, $\log(e^-)$ vs $\log(e^+)$, at three different wavelet scales, $T=35$s, $T=49$s, $T=556$s, and for the same regimes as in figure 3 (top row: all data, second row: $\theta_{Br} > 160 \degree$, third row: $\theta_{Br} < 90\degree$). Dashed lines represent lines of constant positive (lower right) or negative (upper left) cross helicity.}
    \label{fig:elsasser}
\end{figure*}
In addition, the range of wavelet scales over which we see the negative $\nu_c$ population, and the scale at which it disappears, tells us something about the characteristic scale of the switchbacks at $0.17$AU. In these data, switchbacks appear to last on the order of $20-100$s, and their signature has completely disappeared at scales of $\approx 300$s and longer (hence why the bottom right histogram in figure \ref{fig:circles_angle} is empty). This isn't to say switchbacks longer than this never occur. \cite{dudokdewit2019} present evidence that distributions of switchback deflections and residence times are power-law like, so the lack of a signature above 300s in our data set is more likely a finite sampling effect rather than a hard cutoff on the timescales of switchbacks.

Joint probability distributions of $\sigma_r$ and $\nu_c$ have been constructed many times before \citep{bruno2007magnetically,d2010radial,bavassano2006distribution,bavassano1998cross} in a variety of solar wind conditions and heliospheric distances. In particular, \cite{bruno2007magnetically} looked at slow wind using HELIOS 2 data at 0.32, 0.69, and 0.90AU. The features of their plots are broadly similar to ours (they remark there is little radial evolution in slow wind), but there is no sign of a negative cross helicity population similar to what is seen in figure \ref{fig:circles_all}. This is not because switchbacks have disappeared once you are at radial distances of 0.3AU or greater (indeed, they have been directly observed in HELIOS high speed solar wind data prior to PSP, \cite{horbury2018short}) but is a matter of scale. Given that the characteristic timescale of switchbacks at this radial distance of 0.17AU is on the order of tens of seconds, the hour long timescale used in \cite{bruno2007magnetically} is, certainly at smaller heliospheric distances, far too long to observe the switchbacks. It is worth noting that \cite{bruno2007magnetically} do observe a population of negative cross helicity at larger heliospheric distances but due to the large associated negative values of residual energy interpret it as being due to advected structures rather than inward propagating Alfv\'enic waves.
%
%
%
%
%

\begin{figure}
    \centering
    \includegraphics[width=0.48\textwidth]{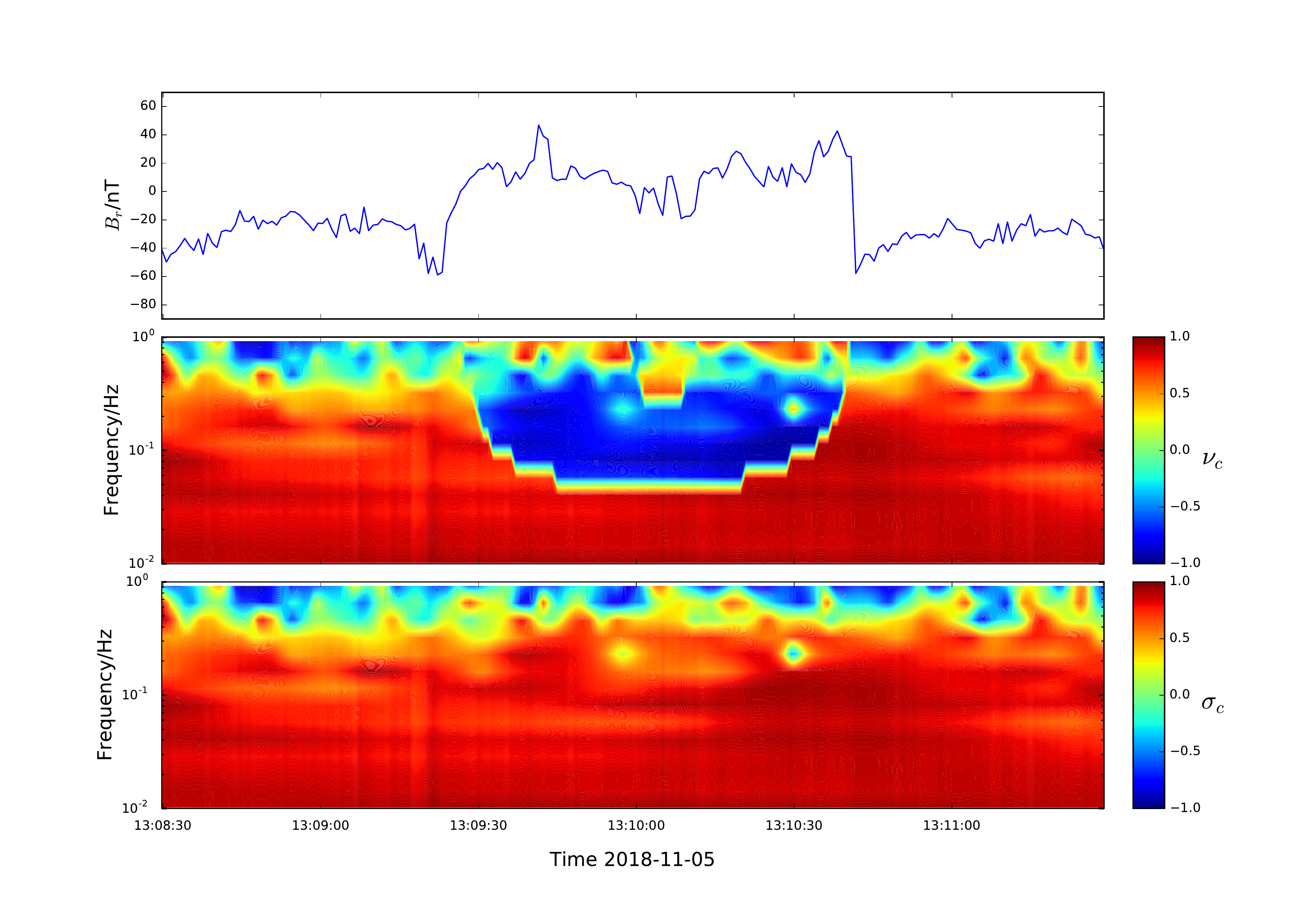}
    \caption{Behaviour of cross helicity through a magnetic switchback. Top panel shows the radial magnetic field. Middle panel shows the wavelet spectrum of rectified cross helicity, $\nu_c(f,t)$, as a function of frequency and time. Bottom panel shows the wavelet spectrum of cross helicity  $\sigma_c(f,t)$.}
    \label{fig:sback_case_study}
\end{figure}

An alternative way of looking at this is shown in figure \ref{fig:elsasser}, where we plot (rectified) Elsasser power, $\log (e^-)$ vs $\log (e^+)$, at the same three wavelet scales and $\theta_{Br}$ regimes as in Figures \ref{fig:circles_all} and \ref{fig:circles_angle}. The diagonal dashed lines represent lines of constant cross helicity $\nu_c$ (from top left to bottom right, $\nu_c = -0.99,-0.8,0.0,0.8,0.99$). Again, the negative cross helicity population is seen at the two shorter scales, but not the longer 560s time scale. Splitting the data up by $\theta_{Br}$ isolates the negative $\nu_c$ population to be due to switchbacks, when $\theta_{Br} < 90\degree$. Both the positive and negative $\nu_c$ distributions are strongly peaked along lines of constant $\nu_c$.

Finally, in Figure \ref{fig:sback_case_study} we show time series of $B_r$, and wavelet spectra of $\nu_c(f,t)$ and $\sigma_c(f,t)$ through a single switchback. The reversal in sign of $\nu_c$ is clearly visible, further supporting the interpretation that the MHD waves are following local field lines at their own scale through the switchback. The region of the spectrogram with negative $\nu_c$ does not extend to all lower frequencies (the ``stepped" appearance of the feature in the $\nu_c$ spectrogram is a visual artifact - it is effectively the cone of influence of the edge-like feature in the magnetic field). At frequencies $ f \lesssim 4\times 10^{-2}$ Hz, the local mean field no longer sees a field reversal because it has been smoothed over a time window that is sufficiently long compared to the time scale of the switchback. Writing $\mathbf{B}(s,t) = \mathbf{B}_0(s,t) + \delta \mathbf{B}(s,t)$, one can think of the switchback as having moved from the local mean field into the fluctuations at some sufficiently large scale, and so $\nu_c = \sigma_c$ at low frequencies.

Regarding the use of $\nu_c$ and $\sigma_c$ effectively as probes of wave propagation direction, of course from Figure \ref{fig:sback_case_study} one can come to the same physical conclusion by examining the behaviour of either $\nu_c$ or $\sigma_c$. One advantage however of $\nu_c$ over $\sigma_c$ is that it gives us statistical information on the characteristic timescales of these events, whereas $\sigma_c$ does not.



%
%

\section{Conclusion}
We have considered the 2D joint distributions of normalised residual energy $\sigma_r(s,t)$ and normalised rectified cross helicity $\nu_c(s,t)$ during one day of PSP's first encounter as a function of scale, $s$. The broad features of the distributions are similar to previous observations in the slow solar wind at small heliocentric distances \citep{bruno2007magnetically}, with highly correlated and Alfv\'enic fluctuations ($\nu_c \sim 0.9$, $\sigma_r \sim -0.3$), but at shorter scales a second population with $\nu_c < 0$ is observed. 

We interpret this to be due to the presence of magnetic switchbacks, and confirm this by splitting the data up according to $\theta_{Br}$, the angle between the scale dependent local mean magnetic field and the radial direction and observing the second population to only appear during switchback times. We conclude that MHD waves are following the \emph{local} magnetic field inside switchbacks, even when it undergoes a large traversal. Predominantly outward propagating flux briefly becomes inward propagating during the field reversal. This also implies that these are local kinks in the magnetic field, and not due to regions of opposite polarity at the Sun's surface. Our analysis provides a useful way to distinguish between these scenarios using only in situ data. $\sigma_c$, as a measure of correlation between $\delta \mathbf{v}_\perp$ and $\delta \mathbf{b}_\perp$ is unaffected by the local mean field direction, showing that the switchbacks are just as Alfv\'enic as the surrounding wind and so switchbacks are in some sense an intrinsic part of it. Propagation direction, as encoded by $\nu_c$, \emph{is} sensitive to the local mean field direction - that is it follows it. This interpretation is further confirmed by directly looking at Elsasser flux inside and outside switchbacks, and a case study following $\nu_c(f,t)$ as a function of time through a single switchback. Computing averaged values of rectified cross helicity without taking into account the reversal effect of switchbacks may result in underestimation of $\nu_c$, an effect which may become more important in later PSP orbits, depending on how the distribution of switchbacks change closer to the Sun.

Finally, a wavelet representation of rectified cross helicity $\nu_c(s,t)$ is seen to be a useful tool for directly observing the inward travelling flux during a large polarity reversing switchback, as well as providing statistical information about the characteristic time scales of switchbacks, which we observe to be in the range 20-100s during this interval. 



\bibliography{biblio}

\begin{thebibliography}{}
\expandafter\ifx\csname natexlab\endcsname\relax\def\natexlab#1{#1}\fi
\providecommand{\url}[1]{\href{#1}{#1}}

\bibitem[{Addison(2017)}]{addison2017illustrated}
Addison, P.~S. 2017, The illustrated wavelet transform handbook: introductory
  theory and applications in science, engineering, medicine and finance (CRC
  press)

\bibitem[{Badman {et~al.}(2019)Badman, Bale, W., {et~al.}}]{badman2019PFSS}
Badman, S.~T., Bale, S., W., B.~J., {et~al.} 2019, The Astrophysical Journal,
  Submitted, this volume ***

\bibitem[{Bale {et~al.}(2016)Bale, Goetz, Harvey, Turin, Bonnell, De~Wit,
  Ergun, MacDowall, Pulupa, Andr{\'e}, {et~al.}}]{bale2016fields}
Bale, S., Goetz, K., Harvey, P., {et~al.} 2016, Space science reviews, 204, 49

\bibitem[{Balogh {et~al.}(1999)Balogh, Forsyth, Lucek, Horbury, \&
  Smith}]{balogh1999heliospheric}
Balogh, A., Forsyth, R., Lucek, E., Horbury, T., \& Smith, E. 1999, Geophysical
  research letters, 26, 631

\bibitem[{Bavassano \& Bruno(2006)}]{bavassano2006distribution}
Bavassano, B., \& Bruno, R. 2006in , 3179--3184

\bibitem[{Bavassano {et~al.}(1998)Bavassano, Pietropaolo, \&
  Bruno}]{bavassano1998cross}
Bavassano, B., Pietropaolo, E., \& Bruno, R. 1998, Journal of Geophysical
  Research: Space Physics, 103, 6521

\bibitem[{Boldyrev(2005)}]{boldyrev2005spectrum}
Boldyrev, S. 2005, The Astrophysical Journal Letters, 626, L37

\bibitem[{Bowen {et~al.}(2018)Bowen, Mallet, Bonnell, \&
  Bale}]{bowen2018impact}
Bowen, T.~A., Mallet, A., Bonnell, J.~W., \& Bale, S.~D. 2018, The
  Astrophysical Journal, 865, 45

\bibitem[{Bruno \& Carbone(2005)}]{bruno2005solar}
Bruno, R., \& Carbone, V. 2005, Living Reviews in Solar Physics, 2, 4

\bibitem[{Bruno {et~al.}(2007)Bruno, d'Amicis, Bavassano, Carbone, \&
  Sorriso-Valvo}]{bruno2007magnetically}
Bruno, R., d'Amicis, R., Bavassano, B., Carbone, V., \& Sorriso-Valvo, L.
  2007in , 1913--1927

\bibitem[{Chandran \& Perez(2019)}]{chandran2019reflection}
Chandran, B.~D., \& Perez, J.~C. 2019, Journal of Plasma Physics, 85

\bibitem[{Chen {et~al.}(2013)Chen, Bale, Salem, \& Maruca}]{chen2013residual}
Chen, C., Bale, S., Salem, C., \& Maruca, B. 2013, The Astrophysical Journal,
  770, 125

\bibitem[{Chen {et~al.}(2011)Chen, Mallet, Yousef, Schekochihin, \&
  Horbury}]{chen2011anisotropy}
Chen, C., Mallet, A., Yousef, T., Schekochihin, A., \& Horbury, T. 2011,
  Monthly Notices of the Royal Astronomical Society, 415, 3219

\bibitem[{Dudok~de Wit {et~al.}(2019)Dudok~de Wit, Bale, W., Bowen,
  {et~al.}}]{dudokdewit2019}
Dudok~de Wit, T., Bale, S.~D., W., B.~J., Bowen, T.~A., {et~al.} 2019, The
  Astrophysical Journal, Submitted, this volume ***

\bibitem[{D’Amicis {et~al.}(2010)D’Amicis, Bruno, Pallocchia, Bavassano,
  Telloni, Carbone, \& Balogh}]{d2010radial}
D’Amicis, R., Bruno, R., Pallocchia, G., {et~al.} 2010, The Astrophysical
  Journal, 717, 474

\bibitem[{Farge(1992)}]{farge1992wavelet}
Farge, M. 1992, Annual review of fluid mechanics, 24, 395

\bibitem[{Fox {et~al.}(2016)Fox, Velli, Bale, Decker, Driesman, Howard, Kasper,
  Kinnison, Kusterer, Lario, {et~al.}}]{fox2016solar}
Fox, N., Velli, M., Bale, S., {et~al.} 2016, Space Science Reviews, 204, 7

\bibitem[{Horbury {et~al.}(2018)Horbury, Matteini, \&
  Stansby}]{horbury2018short}
Horbury, T., Matteini, L., \& Stansby, D. 2018, Monthly Notices of the Royal
  Astronomical Society, 478, 1980

\bibitem[{Horbury {et~al.}(2008)Horbury, Forman, \&
  Oughton}]{horbury2008anisotropic}
Horbury, T.~S., Forman, M., \& Oughton, S. 2008, Physical Review Letters, 101,
  175005

\bibitem[{Karpen {et~al.}(2017)Karpen, DeVore, Antiochos, \&
  Pariat}]{karpen2017reconnection}
Karpen, J., DeVore, C., Antiochos, S., \& Pariat, E. 2017, The Astrophysical
  Journal, 834, 62

\bibitem[{Kasper {et~al.}(2016)Kasper, Abiad, Austin, Balat-Pichelin, Bale,
  Belcher, Berg, Bergner, Berthomier, Bookbinder, {et~al.}}]{kasper2016solar}
Kasper, J.~C., Abiad, R., Austin, G., {et~al.} 2016, Space Science Reviews,
  204, 131

\bibitem[{Livi {et~al.}(2019)Livi, Larson, Kasper, {et~al.}}]{livi2019span}
Livi, R., Larson, D.~E., Kasper, J.~C., {et~al.} 2019, The Astrophysical
  Journal, Submitted, this volume ***

\bibitem[{Mallet \& Schekochihin(2016)}]{mallet2016statistical}
Mallet, A., \& Schekochihin, A.~A. 2016, Monthly Notices of the Royal
  Astronomical Society, 466, 3918

\bibitem[{Neugebauer \& Goldstein(2013)}]{neugebauer2013double}
Neugebauer, M., \& Goldstein, B.~E. 2013in , AIP, 46--49

\bibitem[{Perez \& Chandran(2013)}]{perez2013direct}
Perez, J.~C., \& Chandran, B.~D. 2013, The Astrophysical Journal, 776, 124

\bibitem[{Podesta(2009)}]{podesta2009dependence}
Podesta, J. 2009, The Astrophysical Journal, 698, 986

\bibitem[{Roberts {et~al.}(1987)Roberts, Klein, Goldstein, \&
  Matthaeus}]{roberts1987nature}
Roberts, D., Klein, L., Goldstein, M., \& Matthaeus, W. 1987, Journal of
  Geophysical Research: Space Physics, 92, 11021

\bibitem[{Torrence \& Compo(1998)}]{torrence1998practical}
Torrence, C., \& Compo, G. 1998, A practical guide to wavelet analysis, B. Am.
  Meteorol. Soc., 79, 61--78, ,

\bibitem[{Uritsky {et~al.}(2017)Uritsky, Roberts, DeVore, \&
  Karpen}]{uritsky2017reconnection}
Uritsky, V.~M., Roberts, M.~A., DeVore, C.~R., \& Karpen, J.~T. 2017, The
  Astrophysical Journal, 837, 123

\bibitem[{Velli {et~al.}(1989)Velli, Grappin, \& Mangeney}]{velli1989turbulent}
Velli, M., Grappin, R., \& Mangeney, A. 1989, Physical Review Letters, 63, 1807

\bibitem[{Whittlesey {et~al.}(2019b)Whittlesey, Larson, Kasper, Korreck, Case,
  Livi, {et~al.}}]{whittlesey2019strahl}
Whittlesey, P., Larson, D., Kasper, J.~C., {et~al.} 2019b, The Astrophysical
  Journal, Submitted, this volume ***

\bibitem[{Whittlesey {et~al.}(2019a)Whittlesey, Larson, Jalekas, Abiad,
  {et~al.}}]{whittlesey2019span}
Whittlesey, P., Larson, D.~E., Jalekas, J., Abiad, R., {et~al.} 2019a, The
  Astrophysical Journal, Submitted, this volume ***

\bibitem[{Wicks {et~al.}(2013)Wicks, Roberts, Mallet, Schekochihin, Horbury, \&
  Chen}]{wicks2013correlations}
Wicks, R.~T., Roberts, D.~A., Mallet, A., {et~al.} 2013, The Astrophysical
  Journal, 778, 177

\bibitem[{Yamauchi {et~al.}(2004)Yamauchi, Suess, Steinberg, \&
  Sakurai}]{yamauchi2004differential}
Yamauchi, Y., Suess, S.~T., Steinberg, J.~T., \& Sakurai, T. 2004, Journal of
  Geophysical Research: Space Physics, 109

\end{thebibliography}

\end{document}